# Sculpting Topological Modes on Photonic Chips by Artificial Gauge Fields

Zhiyuan Lin[1], Jian Li[1], Wange Song[1,2*], Xueyun Li[1], Haoran Xin[1], Xian Long[1], Chen Chen[1], Shining Zhu[1], Tao Li[1*], and Shuang Zhang[2,3,4,5,6*]

[1]*National Laboratory of Solid State Microstructures, Key Laboratory of Intelligent Optical Sensing and Manipulations, Jiangsu Key Laboratory of Artificial Functional Materials, School of Physics, College of Engineering and Applied Sciences, Nanjing University, Nanjing, 210093, China.*

[2]*New Cornerstone Science Laboratory, Department of Physics, University of Hong Kong, Hong Kong, China.*

[3]*State Key Laboratory of Optical Quantum Materials, The University of Hong Kong, Hong Kong, China.*

[4]*Department of Electronic and Electrical Engineering, University of Hong Kong, Hong Kong, China.*

[5]*Materials Innovation Institute for Life Sciences and Energy (MILES), HKU-SIRI, Shenzhen, China.*

[6]*Quantum Science Center of Guangdong-Hong Kong-Macao Great Bay Area, 3 Binlang Road, Shenzhen, China.*

[*]*E-mail: songwange@nju.edu.cn, taoli@nju.edu.cn, shuzhang@hku.hk*

## Abstract

Significant efforts have been devoted to manipulating topological states, which often manifest as localized modes at interfaces between distinct topological phases. In this work, we demonstrate a versatile approach to sculpting topological modes (TMs) into any desired shapes by incorporating various artificial gauge fields (AGFs)—including scalar, vector, and imaginary gauge potentials—and leveraging the power of artificial neural networks (ANNs). These AGFs enable precise tuning of the dissipation of the TMs across that of bulk modes, facilitating a transition from localized to fully delocalized states. Moreover, ANNs allow precise engineering of these eigenmodes to achieve tailored profiles of topological states, which remain spectrally isolated within the bandgap and exhibit minimal loss compared to other modes. Our theoretical results are experimentally validated on silicon photonic platforms, demonstrating flexible manipulation of TM profiles. This approach enables the design of topological states with customized properties, offering significant potential for diverse applications in photonics and beyond.

Topological physics has profoundly influenced fields ranging from quantum transport[1-3] to classical wave manipulations[4-11]. A key driver of the advancements has been the exploration of novel topological phases and states beyond the existing framework of topological theory[4,12-17], enabled by the tunable parameters of artificial materials[7,8,18-20]. In topological photonics, for instance, advanced micro-nanofabrication techniques have facilitated tailored geometric designs, unlocking many intriguing physical phenomena, including nonlinear topological solitons[21-25], non-Abelian physics[5,26-30], and non-Hermitian effects[31-33]. Notably, specific geometric modulations can introduce effective "fields", known as artificial gauge fields (AGFs), to neutral particles such as photons. These AGFs govern their dynamics, mimicking the behavior of charged particles under real fields[34-38]. By combining AGFs with Floquet engineering, non-Hermiticity, and topological principles, researchers have uncovered novel nonequilibrium topological phases[33,39-46], including topological triple phase transition[39], Floquet topological insulator[42], and synthetic-dimension physics[43]. Moreover, the versatile tunability of AGFs has driven emerging applications, such as broadband optical coupler[47,48], topological insulator lasers[49], and advanced devices in integrated photonics[50-53].

Despite significant efforts to explore new topological states across various platforms, a common characteristic is their exponential localization at interfaces between topologically distinct systems—a hallmark of the bulk-boundary correspondence[4,12,54,55], protected by non-trivial band structures. However, this localization restricts the use of bulk space, with edge states only occupying a small fraction of the overall space of the topological system. Overcoming the challenge of arbitrarily shaping the mode profiles of topological states without compromising their intrinsic topological properties thus becomes critical[56-58]. Imagine TMs that transcend the localized regions, tailored to any desired shape and spanning the entire available space. Such a capability would not only provide deep insights into fundamental topological physics but also unlock significant promise for advancing topological applications across diverse technological domains.

This work seeks to address these limitations by incorporating various types of AGFs—including scalar, vector and imaginary gauge potentials—to achieve fully delocalized TMs with arbitrary tailored mode profiles, assisted by artificial neural networks (ANNs) (Fig. 1). These gauge fields can be precisely designed to control the energy landscape and the quasienergy band structures of the topological states. Specifically, the dissipation of the TMs can be tuned across that of bulk modes under periodic boundary conditions, facilitating transitions from edge-localized to bulk-delocalized configurations. Furthermore, pre-trained neural networks can predict the optimal coupling configurations to generate TMs with desired shapes while preserving their spectral isolation within the bandgap, ensuring robust topological protections. We experimentally validated this theoretical framework on silicon photonic chips, demonstrating flexible manipulation of topological states and successful eigenmode deformation into diverse target shapes. This approach establishes a powerful paradigm for topological modes manipulation through the combined use of diverse gauge potentials and the ANN-driven design, paving the way for advanced applications on topological photonic chips.

**Photonic waveguide lattice with AGFs.** We begin by considering a one-dimensional (1D) photonic waveguide lattice characterized by nearest-neighbor couplings, denoted as $\kappa_{mn}$, between the $m$-th and $n$-th waveguides. The lattice is subjected to external fields, including real scalar ($\Phi_m$) and vector (**A**) potentials, as well as imaginary potentials ($\gamma_m$) at each site $m$. These potentials are implemented by periodically adjusting the waveguide widths, bending the waveguides, and introducing onsite loss elements, respectively[47,53], as depicted in Fig. 1. These external gauge fields introduce a phase factor to the coupling $\kappa_{mn}$ after the gauge transformation, i.e., $\kappa_{mn}^{\text{AGF}} = \kappa_{mn} \exp(i\mathbf{A} \cdot \mathbf{r}_{mn} + i\Delta\varphi_{mn})$, where $\Delta\varphi_{mn} = -\int_0^z \left[\Phi_m(z') - \Phi_n(z')\right]dz'$ is the local scalar potential difference and $\mathbf{r}_{mn}$ is the position vector from waveguide-$m$ to -$n$, with $m = n \pm 1$. Here the potentials are designed to take sinusoidal forms, $\mathbf{A}(z) = -k_0 A_v \Omega \sin(\Omega z)\mathbf{i}$ and $\Phi_n(z) = (-1)^n k_0 A_s \sin(\Omega z)/2n_0$, where $k_0 = 2\pi n_0/\lambda$ is the wavenumber in the ambient medium, $\Omega$ is the modulation frequency, and $A_v$ ($A_s$) is the modulation amplitude of vector (scalar) potential. A detailed analysis of the gauge transformation and the derivation of artificial scalar and vector gauge potentials are provided in **Supplementary Section S1**.

For simplicity, we focus on incorporating the previously discussed artificial gauge fields into the Su-Schrieffer-Heeger (SSH) model. This model features unit cells, each containing two sublattices, X and Y, with alternating coupling strength, $\kappa_1$ and $\kappa_2$, as illustrated in Fig. 1. Based on the analyses above, the artificial gauge field introduces a phase factor to $\kappa_1$ and $\kappa_2$, i.e., $\kappa_{1(2)}^{AGF} = \kappa_{1(2)} \exp(i\theta_{-(+)})$ and $\theta_{\pm}(z) = G_0 d_{\pm} \sin(\Omega z) \pm \varphi_0 \cos(\Omega z)$, where $d_{\pm}$ is the center-to-center distance between neighboring waveguides, and $G_0 = k_0 A_v \Omega$ and $\varphi_0 = k_0 A_s/\Omega$ are the vector and scalar potential constants. Additionally, the imaginary potentials are implemented in a balanced gain/loss manner, i.e., $\pm\gamma$ for two sublattices. The bulk Hamiltonian can then be succinctly expressed as

$$H(k,z) = \begin{pmatrix} i\gamma & \rho(k,z) \\ \rho^*(k,z) & -i\gamma \end{pmatrix}, \tag{1}$$

where $\rho = \kappa_1 e^{i\theta_-} + \kappa_2 e^{-i\theta_+} e^{-ik}$. Since this Hamiltonian is periodic, the topological nature of the system can be well described by Floquet theory, which dictates the quasienergy and Floquet states in our non-Hermitian systems[59].

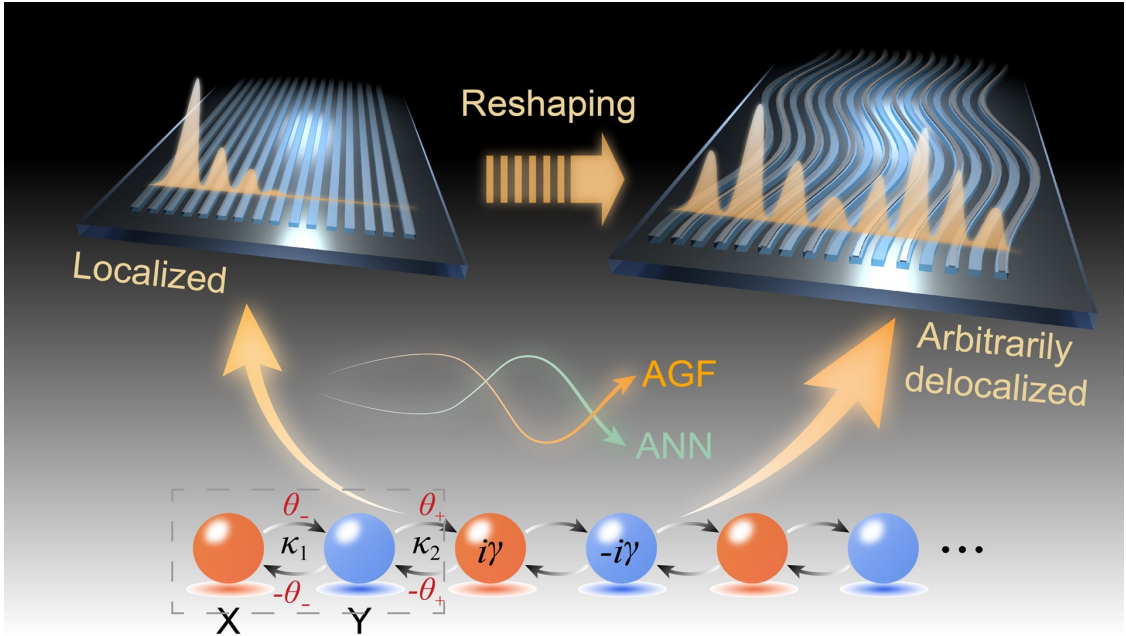

**Fig. 1 | Sculpting topological modes on photonic chips.** The localized topological modes can be arbitrarily reshaped into the desired delocalized shapes with the ANNs and AGFs. The AGFs (combined by the scalar, vector, and imaginary gauge potentials) can be achieved by varying the width, bending trajectory, and adding onsite gain/loss of the waveguides.

**Topological mode control by AGFs.** In the following analysis, we fix the imaginary potentials to $\gamma = \kappa_1$. For $G_0 = \varphi_0 = 0$, the system exhibits parity-time (PT) symmetry, which results in TMs localized at the boundary for $\kappa_1 < \kappa_2$. When $G_0 \neq 0$ and $\varphi_0 \neq 0$, the introduction of AGFs significantly influences the distribution of the TMs. Specifically, transitions of TMs from localized to delocalized states can be observed, which is characterized by the modified inverse participation ratios (MIPR) $\mathcal{M} = \sum_{i \in T_L, j \in T_R} \left( |\langle i | \phi_{\text{TM}} \rangle|^4 - |\langle j | \phi_{\text{TM}} \rangle|^4 \right) \Big/ \left( \sum_{k \in T} |\langle k | \phi_{\text{TM}} \rangle|^2 \right)^2$, where $|\phi_{\text{TM}}\rangle$ is the topological mode, $T_L$ ($T_R$) denotes the left (right) half of the lattice, and $T = T_L \bigcup T_R$. Figure 2a illustrates the phase diagram plotted as functions of scalar and vector potentials, which is separated into two regions by the critical line (black curve, $\mathcal{M} = 0$). The regions with $\mathcal{M} > 0$ and $\mathcal{M} < 0$ correspond to TMs localized at the left and right boundary, respectively, while the critical line indicates fully delocalized TMs.

Specifically, we investigate the evolution of TMs along the parameter path $l$ indicated by the dashed line in Fig. 2a, starting from the origin. When $G_0 = \varphi_0 = 0$ (pink circle in Fig. 2a), corresponding to the absence of gauge field modulation, the TM lies in the $\mathcal{M} > 0$ region and is localized at the left boundary (pink profile in Fig. 2b). As the parameters deviate from the origin but remain within $\mathcal{M} > 0$ (e.g., at the blue pentagram or purple hexagon in Fig. 2a), the TM begins to extend slightly from the left edge (blue and purple profiles in Fig. 2b). Remarkably, when the parameters reach the critical boundary (upper and inverted triangles in Fig. 2a), the TMs become fully delocalized (green and brown profiles in Fig. 2b). Upon entering the $\mathcal{M} < 0$ region (orange circle in Fig. 2a), the TM shifts its localization to the right boundary (orange profiles in Fig. 2b).

**Mechanism of topological mode reshaping.**

The transition mechanism can be understood through the topological properties of the complex Floquet quasienergy spectrum (Fig. 2c,d). Notably, the quasienergy spectra under periodic boundary conditions (PBC) and open boundary conditions (OBC) differ, signaling the emergence of non-Hermitian skin effects (NHSEs)[10,11,60-63]. Here, the interaction between the NHSEs and TMs gives rise to a hybrid skin-topological effect[9,60], which modifies the mode profiles of the topological states. Specifically, along path $l$, the TM approaches the boundary of the PBC spectrum, traverses into its interior, and subsequently exits it (Fig. 2d). When the TM aligns with the PBC energy spectrum, an equilibrium forms between the NHSE and the opposing localization tendency of the TM, resulting in a fully delocalized mode.

Thus, our design demonstrates flexible generation and manipulation of NHSEs using AGFs, in contrast to previous approaches that rely on nonreciprocal couplings[9-11,60-63], which are difficult to implement in on-chip optical systems. In **Supplementary Section S2,** we rigorously prove the emergence of the NHSE via AGF from a symmetry perspective. Our analysis reveals that both scalar and vector potentials are essential for enabling the transformation of TMs. If either $G_0 = 0$ or $\varphi_0 = 0$, symmetry constraints inhibit the NHSE, thereby preventing the existence of extended TMs. Notably, in the region where $\mathcal{M} < 0$, sufficiently large vector gauge potentials can drive the system into a pure skin mode regime (enclosed by the white curve), in which TMs disappear and merge into conventional skin modes. By tracing the behavior along parameter path $l$ as shown in Fig.2, the transformation of topological mode profiles from localized to delocalized states is clearly illustrated. A more detailed analysis of the transition process across the skin mode regions—including comprehensive quasienergy spectra and physical interpretations—is provided in **Supplementary Section S3**.

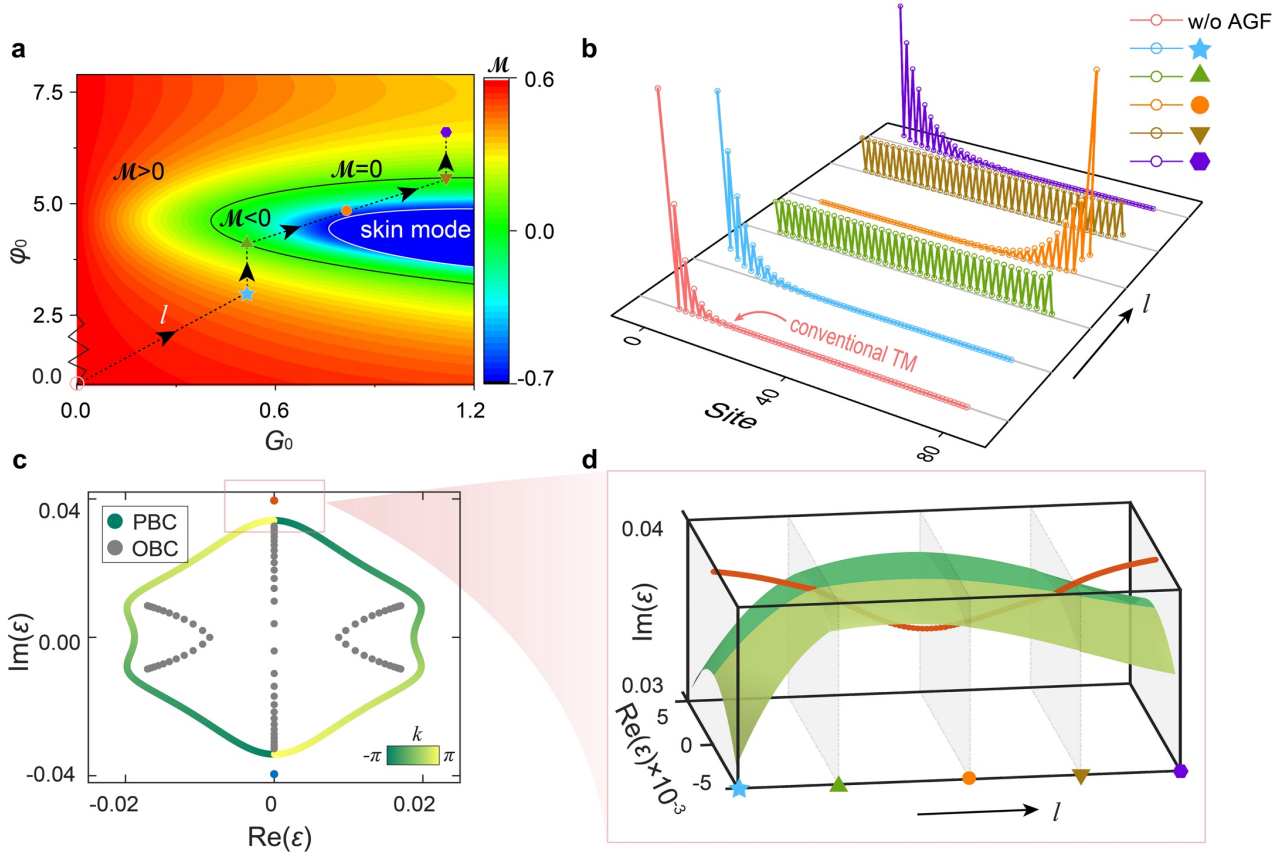

**Fig. 2 | Topological transformation via scalar and vector gauge potentials. a**, A phase diagram illustrating the distribution of topological states as functions of the scalar potential ($\varphi_0$) and the vector potential ($G_0$). In the parameter path $l$, six sets of parameters are selected to describe the reshaping process, i.e., (0, 0), (0.5, 2.96), (0.5, 4.13), (0.8, 4.87), (1.1, 5.53), and (1.1, 6.58). **b**, The corresponding profile of topological mode under the selected parameter in **a**. **c**, The quasienergy spectrum of the parameter marked by blue pentagrams in **a**. Here the red and blue dots represent the topological modes. **d**, The evolution of topological modes and quasienergy spectra along the parametric path $l$. The other parameters are chosen as $\kappa_1 = 0.01k_0$, $\kappa_2 = 2\kappa_1$, $\gamma = \kappa_1$, $\Omega = 3\kappa_2$ and the site number $N = 80$.

**Neural networks-assisted arbitrary topological sculpting.**

The phase diagram shown in Fig. 2a demonstrates that by jointly modulating the scalar and vector gauge potentials, a rich variety of phase transitions can be realized, including localized TMs, delocalized TMs, and skin modes. It is noteworthy that the delocalized TMs inherently exhibit a uniform distribution, as shown in Fig. 2b (green mode). This distribution can be further tailored into more complex profiles—such as sinusoidal ("Type I") and convex ("Type II") forms, illustrated in Fig. 3a—by utilizing artificial neural networks (ANNs) to implement a non-uniform distribution of coupling coefficients, thereby controlling the spatial distribution of AGFs. This approach involves three key steps: first, generating a large dataset (18,000 sets in this work) by calculating the topological modes of the Floquet Hamiltonian for various coupling coefficients, while keeping all other parameters the same as those of the initial uniform delocalized TMs. Second, the ANNs are trained on this dataset, mapping normalized inputs to outputs by optimizing the hidden-layer weights via backpropagation algorithm (Fig. 3b). Finally, the desired TMs are provided as inputs to the trained ANNs to predict the required coupling coefficients, which are then used to construct the predicted TM—evaluated using mean square error (MSE)—and these coefficients correspond to the physical waveguide parameters needed to realize the target structure (Fig. 3c). More details regarding the training process and the prediction results of the neural networks can be found in **Supplementary Section S4**.

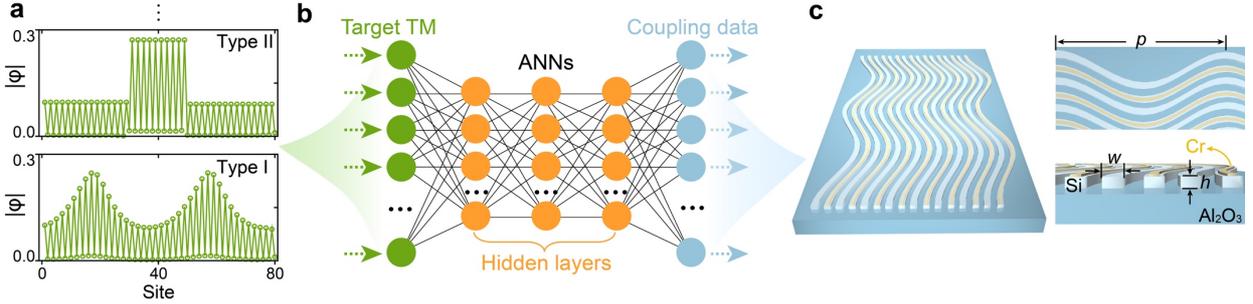

**Fig. 3 | Arbitrary sculpting of topological modes using an artificial neural network. a**, The desire topological modes, e.g., sinusoidal type ("Type I") and convex type ("Type II"). **b**, The amplitude of the topological mode is used as an input layer, which is passed through a hidden layer and finally outputs the desired coupling data, which will correspond to the actual waveguide parameters to finally obtain the desired waveguide structure in **c**.

**Experiments**

**Manipulation of NHSEs by gauge potentials.** The configurations can be conveniently implemented on an integrated photonic platform, where scalar and vector potentials are achieved by varying the width and bending trajectory of the waveguides, respectively. In the experiments, we first verified that the interaction of different types of gauge fields can lead to NHSEs in silicon photonics. The optical lattice sample comprised of $N = 13$ Si waveguides on the sapphire substrate with air cladding. The waveguide height is $h = 220$ nm and the waveguide spacing is $g = 200$ nm. The trajectory and width of the waveguide are periodically modulated in a sinusoidal way, with modulation period $p = 25$ μm, the bending amplitude $A_v \approx 1$ μm, and the width $w$ variation interval around $360 \sim 420$ nm. The non-Hermiticity is introduced by coating a layer of chromium (Cr) with width $w_c = 200$ nm and thickness $h_c = 4$ nm on top of every other Si waveguide. The scanning electron microscope (SEM) pictures of experimentally fabricated samples are shown in Fig. 4a, where the waveguide structure and deposited Cr can be clearly observed (see **Methods** for fabrication details).

To demonstrate the feasibility of this scheme, two samples with opposite scalar potentials ($\varphi_0 = \pm 4.1$) were designed, featuring reversed arrangements of neighboring waveguide widths (Fig. 4a). These two configurations correspond to distinct winding topologies, $\mathcal{W} = \pm 1$ (see **Supplementary Section S2**), and are thus expected to display opposite evolution trends. Each type of sample was fabricated with varying lengths ($L = p, 2p, 3p,$ and $4p$) to capture various stages of mode evolution. In optical measurements, the center of the waveguide lattices is excited by a near-infrared laser with wavelength $\lambda = 1550$ nm via a grating coupler, and the output light scattered from the end of the waveguides was measured by a near-infrared camera through a microscope objective (see **Methods** for measurement details).

The experimental results and the simulated light evolutions (using the commercial finite-element software COMSOL Multiphysics) are presented in Fig. 4b-4e, where Fig. 4c and 4e are the extracted data for further verification. For the samples with winding number $\mathcal{W} = -1$, light in the lattice gradually evolves towards the right end of the boundary, as shown in Fig. 4b and 4c. In contrast, when $\mathcal{W} = +1$, light tends to evolve toward the opposite boundary (Fig. 4d and 4e). These results confirm our theoretical predictions and lay the groundwork for subsequent reshaping of TMs in waveguide lattices.

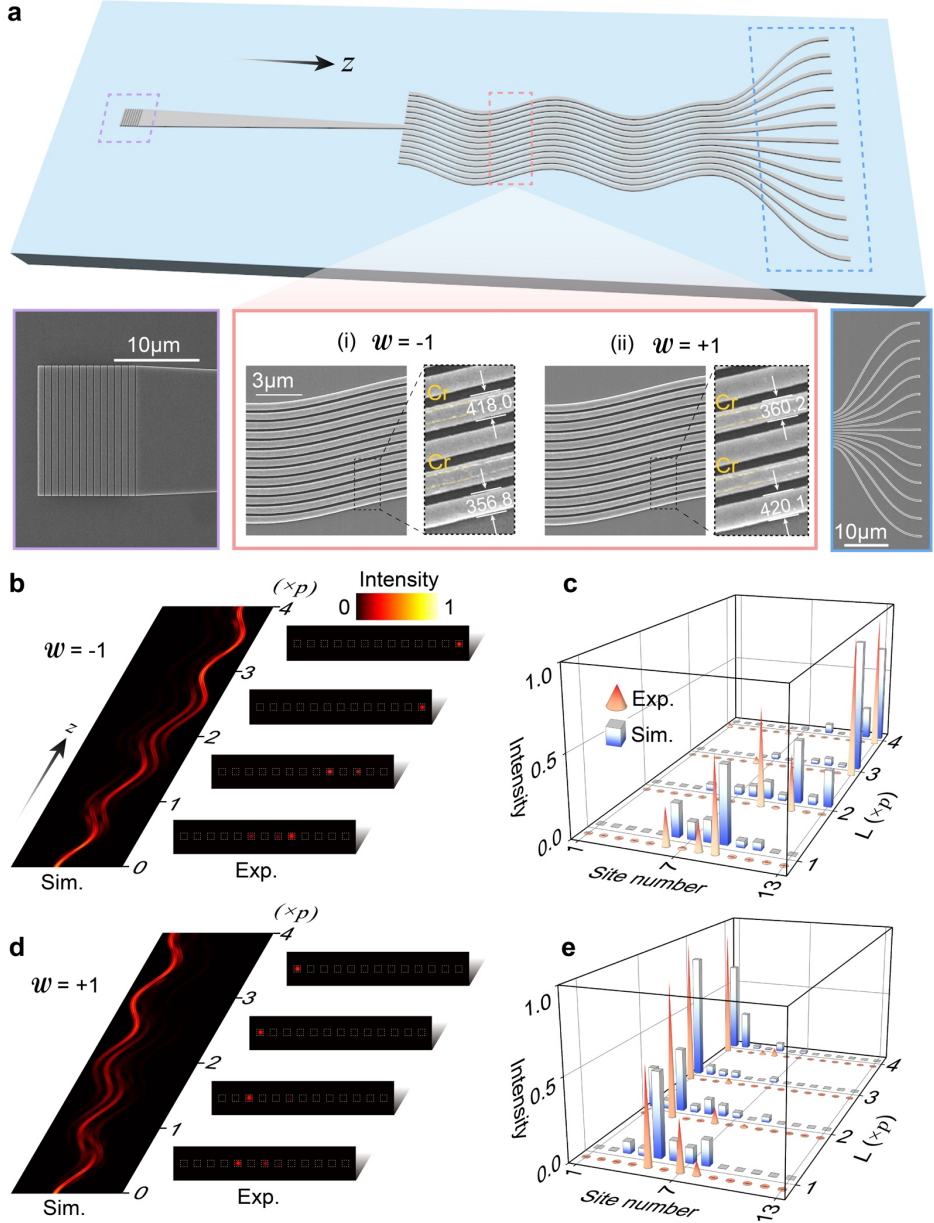

**Fig. 4 | Manipulation of NHSEs in silicon photonics by gauge potentials. a**, The schematic of the waveguide structure and the enlarged SEM pictures, where the deposited Cr strips and the width difference of the two topological designs, (i) $\mathcal{W} = -1$ and (ii) $\mathcal{W} = +1$ can be clearly observed. **b-e**, Simulation results and the experimentally measured light intensities for light propagation along the $z$ direction at different propagation lengths ($L = p, 2p, 3p$, and $4p$, $p = 25$ μm), where **c** and **e** are the extracted data. The intensity is normalized to 1 at every $z$ in the simulation results.

**Observation of photonic topological delocalized modes.** To obtain extended TMs predicted in Fig. 2, we fabricated waveguide samples ($N = 16$) with alternate spacing ($g = 200$ nm and 300 nm) to introduce the interaction of TM with NHSE. Such an arrangement supports a TM with minimal loss localized at the left boundary. Next, gauge fields were introduced to drive skin modes toward opposite boundary (similar to NHSE with $\mathcal{W} = -1$ in Fig. 4b) to obtain a fully delocalized TM, where the bending amplitude is chosen as $A_v \approx 0.5$ μm, the remaining parameters are consistent with Fig 4b.

The excitation of these delocalized states requires careful design compared to localized states. The fully delocalized TM, primarily distributed on Cr-free waveguides with uniform intensity but anti-phase between neighboring sites, necessitates a cascaded array for initial state preparation. Here, light coupled from an input grating passes through a Y-branch splitter and then through a waveguide incorporating subwavelength grating (SWG) microstructures to generate eight anti-phase outputs (Fig. 5a). This initial state can successfully excite a fully delocalized topological eigenmode in the array, preserving its shape during propagation. The simulation and experimental results of its output are in good agreement, as shown in Fig. 5c. However, when the topological delocalized state is not an eigenmode of the system, for instance, by setting the scalar potential to zero ($\varphi_0 = 0$), the topological eigenmode with minimal loss becomes localized. Consequently, the initial state evolves towards the boundary occupied by the TM, as shown in Fig. 5b. These comparative results confirm the successful realization of fully delocalized topological eigenmodes in the waveguide lattice.

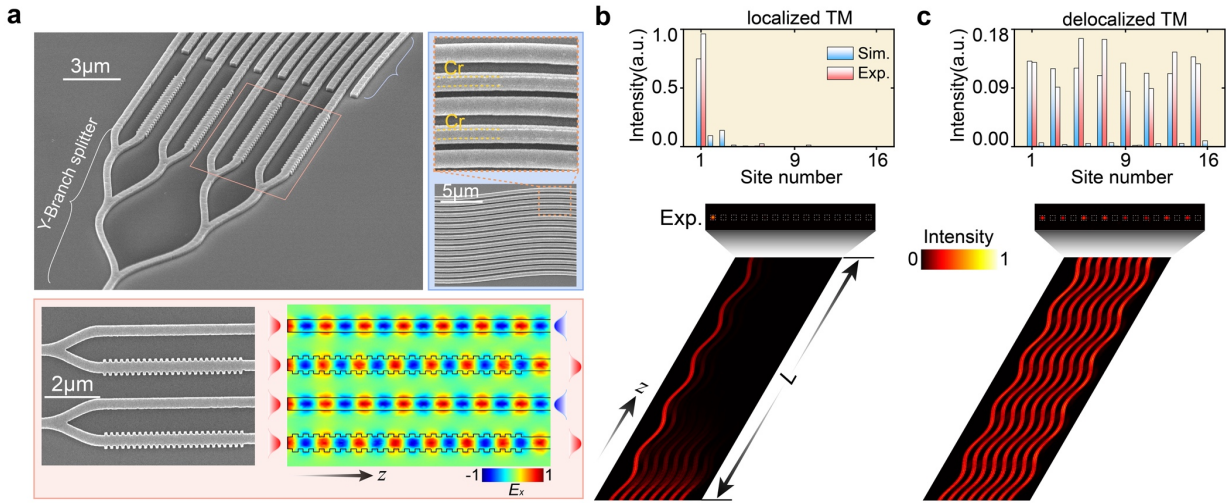

**Fig. 5 | Observation of photonic topological delocalized modes. a**, Schematic of the silicon waveguide structure, wherein the initial state preparation comprises the Y-branch splitter and the waveguide incorporated with the SWG microstructure to generate anti-phase outputs. Simulation results and the experimentally measured light intensities of the waveguide lattices supporting topological **b**, localized and **c**, delocalized modes, where the propagation distance is $L = 5p$.

**Observation of topological modes incorporating diverse shapes.** With the assistance of ANNs, we can efficiently determine the suitable parameters for achieving the desired shapes of topological states. In the experiments, we fine-tuned the waveguide spacing ($g_1, g_2, g_3, \ldots, g_N$) to match the distribution of coupling coefficients obtained from the networks. We conducted a series of experiments to verify two types of extended topological states, the sinusoidal type (Type I) and the convex type (Type II) (Fig. 3). These states exhibit distinct spatial distributions and require precise initial conditions for proper excitation. Exciting these complex extended states requires not only anti-phase initial conditions but also a non-uniform intensity distribution across the waveguides. Achieving such initial conditions is challenging but feasible through an inverse design method, which systematically optimizes input parameters to produce the desired output (Fig. 6a). Finally, experimental results for both Type I and Type II extended states were obtained and compared with simulation results (Fig. 6b and 6c). The results demonstrate good agreement between the experimental observations and the theoretical predictions, confirming the effectiveness of our approach.

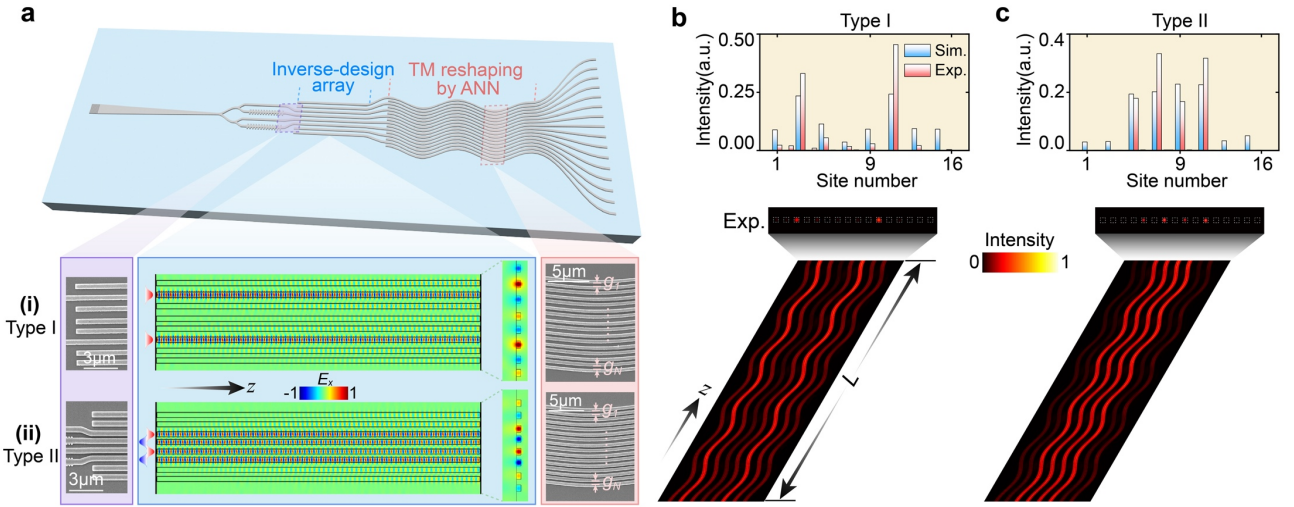

**Fig. 6 | Demonstration of diverse shapes of photonic topological modes. a**, Schematic of the waveguide design with inverse-designed array for initial states and the artificial neural network-assisted waveguide array (top panel). The inverse-designed array for (i) Type I: sinusoidal type and (ii) Type II: convex type (bottom panel). **b**, **c**, The simulation and experimental results for Type I and Type II lattices, respectively.

We have demonstrated customized sculpting of topological states using artificial gauge fields,

achieved by modulating scalar, vector, and imaginary gauge potentials on a silicon photonic chip. Our work also establishes a critical synergy between topological photonics and artificial neural network design, enabling unprecedented control over quasienergy band structures. This approach facilitates the tuning of topological modes across the boundary of bulk states and transitions from localized to fully delocalized states with well-defined mode profiles—empowered by artificial neural networks. Future research holds significant promise: for example, our design paradigm can be extended to higher-dimensional systems, unlocking richer gauge field configurations and paving the way for customized topological states with diverse applications. Moreover, our results highlight the potential of artificial neural networks to efficiently optimize the gauge field configurations tailored to specific photonic functionalities. For example, these advances could drive the realization of topological broad-area single-mode laser with arbitrarily shaped wavefront, and enhance optical communications and signal processing, where precise manipulation of light propagation and mode structure is critical.